\documentclass{ieeeaccess}

\usepackage[hyphens]{url}
\usepackage[hidelinks]{hyperref}
\hypersetup{breaklinks=true}
\usepackage{cite}
\usepackage{amsmath,amssymb,amsfonts}
\usepackage{algorithmic}
\usepackage{graphicx}
\usepackage{textcomp}
\usepackage{afterpage}
\usepackage{array}
\usepackage{supertabular}
\usepackage{multirow}
\usepackage{multicol}
\usepackage{booktabs}

\def\BibTeX{{\rm B\kern-.05em{\sc i\kern-.025em b}\kern-.08em
    T\kern-.1667em\lower.7ex\hbox{E}\kern-.125emX}}

\begin{document}
\history{Date of publication xxxx 00, 0000, date of current version xxxx 00, 0000.}
\doi{10.1109/ACCESS.2021.DOI}

\title{Enhanced Accuracy Simulator for a Future Korean Nationwide eLoran System}
\author{\uppercase{Joon Hyo Rhee\authorrefmark{1}, 
Sanghyun Kim\authorrefmark{2}, 
Pyo-Woong Son\authorrefmark{3},
and Jiwon Seo\authorrefmark{2}} \IEEEmembership{Member, IEEE}}
\address[1]{Korea Research Institute of Standards and Science, Daejeon 34113, Republic of Korea}
\address[2]{School of Integrated Technology, Yonsei University, Incheon 21983, Republic of Korea}
\address[3]{Korea Research Institute of Ships and Ocean Engineering, Daejeon 34103, Republic of Korea}
\tfootnote{This research was a part of the project titled “Development of integrated R-Mode navigation system” funded by the Ministry of Oceans and Fisheries, Korea. Joon Hyo Rhee and Sanghyun Kim are co-first authors.}

\markboth
{Rhee \headeretal: Enhanced Accuracy Simulator for a Future Korean Nationwide eLoran System}
{Rhee \headeretal: Enhanced Accuracy Simulator for a Future Korean Nationwide eLoran System}

\corresp{Corresponding author: Jiwon Seo (jiwon.seo@yonsei.ac.kr)}

\begin{abstract}
The Global Positioning System (GPS) has become the most widely used positioning, navigation, and timing system. However, the vulnerability of GPS to radio frequency interference has attracted significant attention. After experiencing several incidents of intentional high-power GPS jamming trials by North Korea, South Korea decided to deploy the enhanced long-range navigation (eLoran) system, which is a high-power terrestrial radio-navigation system that can complement GPS. As the first phase of the South Korean eLoran program, an eLoran testbed system was recently developed and declared operational on June 1, 2021. Once its operational performance is determined to be satisfactory, South Korea plans to move to the second phase of the program, which is a nationwide eLoran system. For the optimal deployment of additional eLoran transmitters in a nationwide system, it is necessary to properly simulate the expected positioning accuracy of the said future system. In this study, we propose enhanced eLoran accuracy simulation methods based on a land cover map and transmitter jitter estimation. Using actual measurements over the country, the simulation accuracy of the proposed methods was confirmed to be approximately 10\%--91\% better than that of the existing Loran (i.e., Loran-C and eLoran) positioning accuracy simulators depending on the test locations. 
\end{abstract}


\titlepgskip=-15pt

\maketitle

\section{Introduction}
\label{sec:introduction}
\PARstart{T}{he} global positioning system (GPS) is the most widely used positioning, navigation, and timing (PNT) system. Ground vehicles, airplanes, and ships can use GPS signals to obtain their positions, speeds, and heading information. In addition, GPS is used for precise timing applications. Considering the wide applications of global navigation satellite systems (GNSSs), including the GPS system of the United States (US) and Galileo of Europe, the vulnerabilities of GNSSs have become a serious problem. The transmission power of the GNSS signals from satellites is very limited, and the received signal level at user receivers is even lower than the thermal noise level \cite{Braasch99}. Thus, high-power radio frequency interference (RFI) at the appropriate frequency bands can easily disrupt GNSS signals. In addition to RFI, ionospheric anomalies can disrupt GNSS signals \cite{Jiao15, Seo11:Availability, Seo14:Future, Lee17:Monitoring, Sun20:Performance}.

There have been considerable cases in which intentional or unintentional GPS interference have been reported \cite{Rooker:Satell, Carroll08, FAA:GPS, Curry:Sentinel, OSCE:Emergency, Gertz:Air}. Various methods have been studied to mitigate the impact of RFI \cite{DeLorenzo12, Kim19:Mitigation, Chen10, Chen11, Shin17:Autonomous, Rhee19:Low}. A controlled reception pattern antenna (CRPA) \cite{Chen12, Park21:Single, Park18:Dual, Seo11:Real-time} is known to be one of the most effective ways to mitigate RFI. However, its relatively large size and high cost limit its applicability. Furthermore, even a GNSS CRPA may not provide a reliable PNT information under the intentional high-power jamming that South Korea has experienced. South Korea has experienced repeated intentional high-power GPS jamming from North Korea since 2010. For example, during a 6-day period of jamming in 2016, 1,794 cell towers, 1,007 airplanes, and 715 ships experienced GPS disruptions \cite{Son18:Novel}. Considering the numerous critical infrastructures that are dependent on GPS, South Korea decided to deploy an enhanced long-range navigation (eLoran) \cite{Pelgrum06, Williams13, Son19:Universal, Li20} system as a complementary PNT system to GNSSs \cite{Seo13:eLoran}. 

eLoran is a high-power terrestrial radio navigation system whose performance is greatly improved from Loran-C \cite{ILA07:eLoran, Montgomery08, Pu21, Yuan20, Qiu10}. It is virtually impossible to jam eLoran signals because the transmitting power of an eLoran signal is high enough to resist most practical jamming signals. Thus, eLoran can still provide PNT services where GNSSs are unavailable. The United Kingdom (UK) demonstrated the functionality of a prototype eLoran system in October 2014. This particular prototype had better than 10 m accuracy for maritime users \cite{Offermans15} after additional secondary factor (ASF) correction by a differential correction station and ASF maps \cite{Hargreaves12, Williams00:Mapping, Hwang18:TDOA}. The ASF is a signal propagation delay due to the land path, and it is the largest error source for Loran systems (i.e., Loran-C and eLoran) \cite{Lo03, Lo09, Zhou13}. 
After four years of research and development, South Korea has recently deployed an eLoran testbed system \cite{Son20}, which has been operational since June 1, 2021. A new eLoran transmitter was installed in the testbed, and two existing Loran-C transmitters were upgraded. In addition, two differential correction stations were deployed to cover two harbors in the testbed \cite{Son20, Son18:Development}. If the demonstrated performance of eLoran in the testbed is determined to be satisfactory, South Korea will consider using the eLoran system nationwide. To deploy such a nationwide eLoran system, it is necessary to perform performance simulations corresponding to the locations of the additional eLoran transmitters. In particular, the expected positioning accuracy in the coverage area is important because South Korea requires a minimum accuracy of 20 m for eLoran to be used as a maritime backup navigation system, according to the request for proposal of the eLoran testbed project. 

\begin{figure}[t!]
  \centering
  \includegraphics[width=0.6\linewidth]{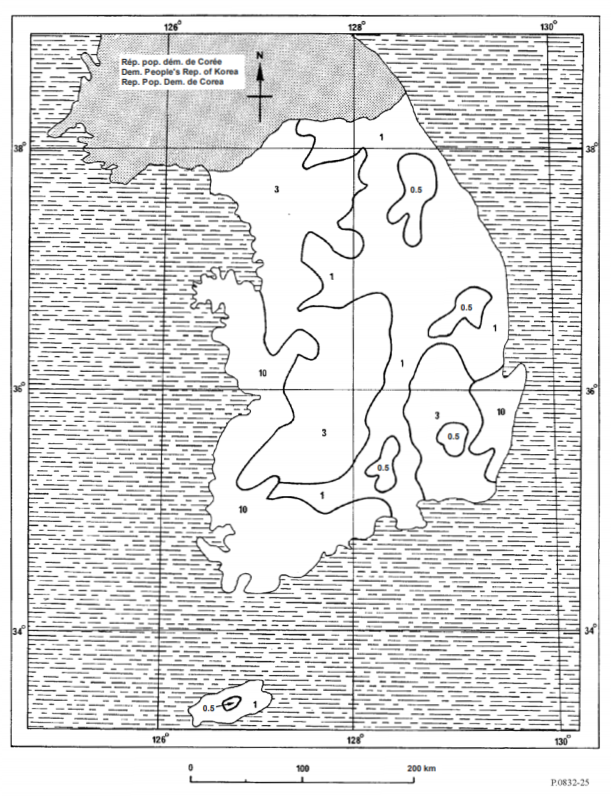}
  \caption{Effective ground conductivity data over Korea (Fig. 25 of \cite{ITU92:World}).}
  \label{fig:EffGndCond}
\end{figure}

There are existing Loran performance simulators that were developed based on the studies in the US \cite{Lo08} and the UK \cite{Hargreaves15}. To estimate the Loran positioning accuracy, the signal-to-noise ratio (SNR) of the received signal and the jitter of each Loran transmitter need to be specified. Previous studies used the effective ground conductivity data from the International Telecommunication Union (ITU) \cite{ITU92:World} to estimate the received signal strength, which is necessary in estimating the SNR. 
However, the effective ground conductivity over Korea from the ITU document shown in Fig.~\ref{fig:EffGndCond} is too coarse for use in a precise simulation. The transmitter jitter is more difficult to specify because each transmitter has different jitter characteristics. Previous studies assumed a fixed jitter (e.g., 6 m \cite{Lo08} or 4 m \cite{Hargreaves15}) for all transmitters for simplicity, which results in additional errors in a positioning accuracy simulation. 
Because of the aforementioned issues, the South Korean government realized the necessity of developing a more accurate Loran accuracy simulator for Korea.

In this study, we propose the use of a land cover map to estimate the received signal strength more accurately than in previously defined methods. The land cover map contains information on the land usage, which can be converted to the effective ground conductivity. Regarding the transmitter jitter, we propose a method to estimate the Loran transmitter jitter based on Loran time-of-reception (TOR) \cite{Son18:Novel} measurements. 
 
The contributions of this work are summarized as follows:
\begin{itemize}
  \item A method for generating fine-grid ground conductivity data based on a land cover map was proposed. We obtained approximately 1,600 ground conductivity values over South Korea using the proposed method, which is far more than the 11 values from the ITU data shown in Fig. \ref{fig:EffGndCond}. The field tests showed approximately 5\%--72\% improvements in the received signal strength simulation compared to the existing method.
  \item A method for estimating the Loran transmitter jitter was proposed. The jitters of the four Loran transmitters in Northeast Asia were estimated using the proposed method. Using the estimated jitters, approximately 9\%--96\% improvements in the measurement error simulation over the existing methods were demonstrated through field tests.
  \item The simulation performance of the Loran positioning accuracy was tested at seven locations in Korea. Overall, our methods of utilizing the land cover map and estimating the transmitter jitters resulted in  approximately 10\%--91\% improvements in estimating the Loran positioning accuracy in Korea over the existing methods. 
\end{itemize}

The remainder of this article is structured as follows. The Loran accuracy simulation method is introduced in Section \ref{sec:LoranAccSim}. The proposed methods for improving the simulation accuracy are presented in Section \ref{sec:ProposedMethods}. After discussing the field test results in Section \ref{sec:ExpValidation}, the conclusions are presented in Section \ref{sec:Conclusion}.

\section{LORAN ACCURACY SIMULATION}
\label{sec:LoranAccSim}
\begin{figure}[t!]
  \centering
  \includegraphics[width=1.0\linewidth]{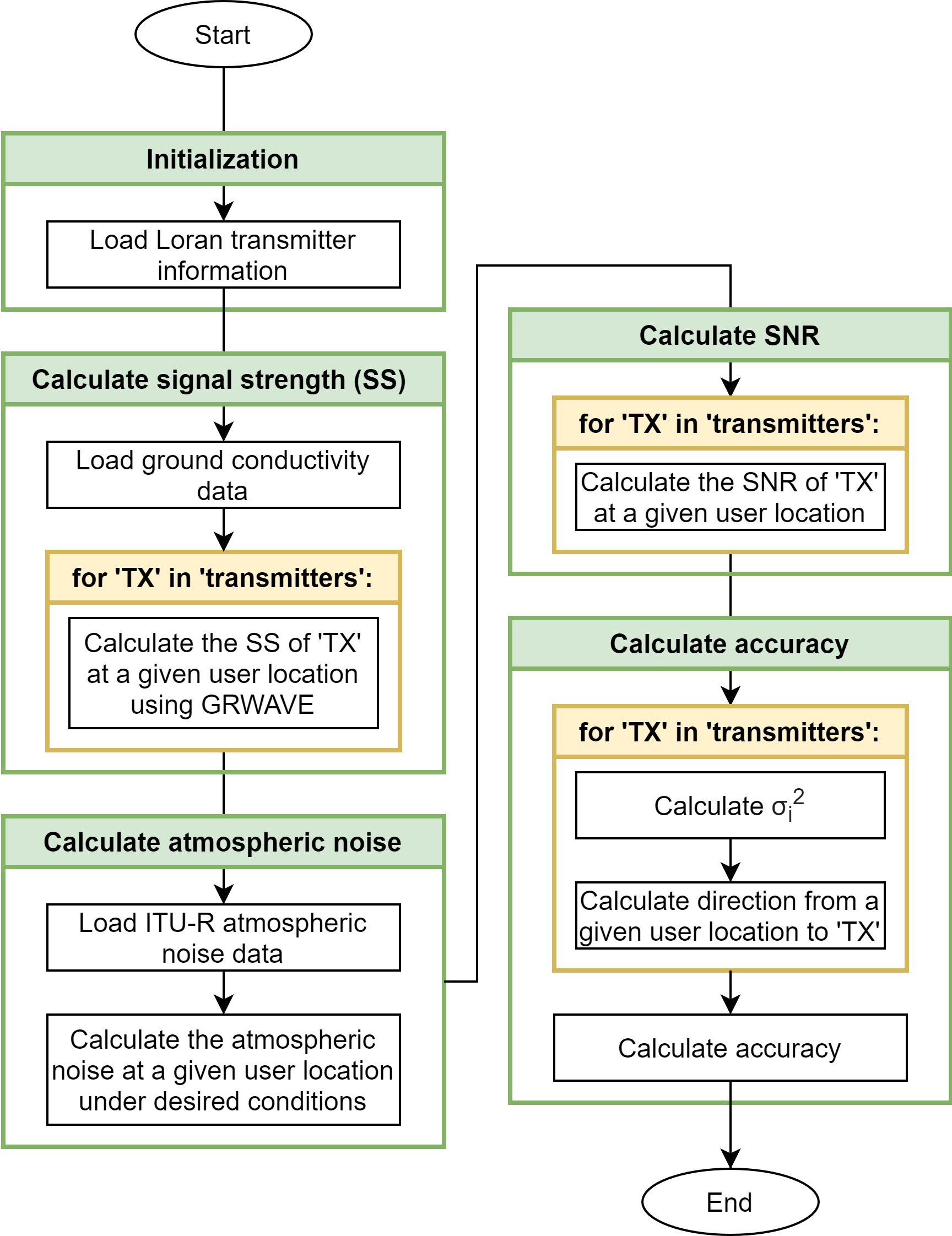}
  \caption{Flowchart of the Loran accuracy simulation procedure.}
  \label{fig:FlowchartAccSim}
\end{figure}

A flowchart of the generic Loran accuracy simulation method is shown in Fig.~\ref{fig:FlowchartAccSim}. Each component of the flowchart is explained in the following subsections.

\subsection{SNR Simulation}
\label{sec:SnrSim}
SNR is the ratio between the received signal strength and the noise strength. Therefore, these two factors must be determined to estimate the SNR. The signal strength diminishes as the signal propagates; thus, it is related to the propagation path. Signal strength decreases as the propagation distance increases, and the decreasing slope is related to physical characteristics, such as frequency, ground conductivity, and permittivity. Rotheram \cite{Rotheram81:Part1,Rotheram81:Part2} presented methods to calculate the expected signal strength, which was reflected in the ITU-R P.368-9 \cite{ITU07:Ground-wave}, and an algorithm was formed into usable software by ITU, namely GRWAVE \cite{Garcia:Calculation},  which was utilized in our work. To calculate the estimated signal strength using GRWAVE, it is necessary to know the effective radiated power (ERP); frequency of the signal; ground conductivity, permittivity, and elevation profile of the propagation path; and the gain and height of the transmitting antenna. Using these data, GRWAVE calculates the expected signal strength in dB($\mathrm{\mu}$V/m).

\begin{figure}[t!]
  \centering
  \includegraphics[width=1.0\linewidth]{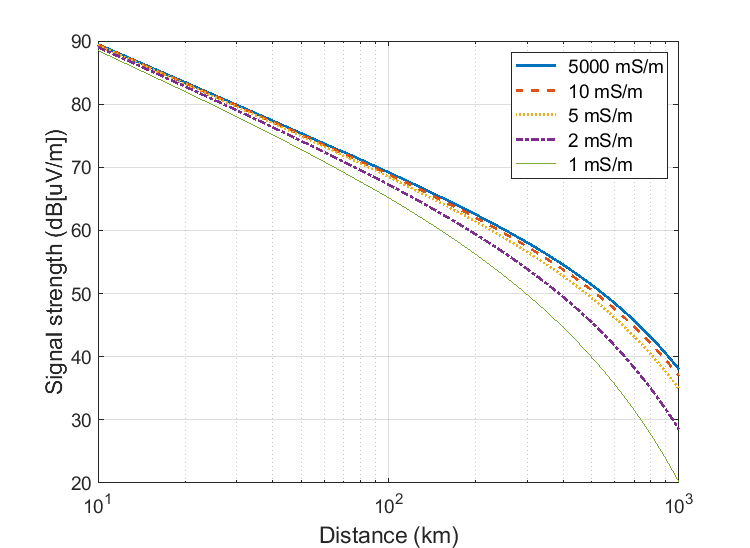}
  \caption{Signal strength in dB($\mathrm{\mu}$V/m) versus distance graph for various effective ground conductivity values, simulated using GRWAVE (based on the data from \cite{Barringer99:Radio}).}
  \label{fig:SigStrength}
\end{figure}

Among these input parameters, ground conductivity requires the most attention because it is more difficult to specify than the other parameters. The ground conductivity along a propagation path directly affects to the signal-strength attenuation. For example, a higher ground conductivity at sea causes less signal attenuation than a lower ground conductivity among mountains. Fig.~\ref{fig:SigStrength} represents an example of the relationship among  signal strength, propagation distance, and effective ground conductivity, simulated using GRWAVE.  

Previous Loran performance simulators \cite{Lo08,Rhee13,Hargreaves15} used the ground conductivity data published by the ITU \cite{ITU92:World}. This ITU document includes the ground conductivity values of the entire world. However, the ground conductivity over Korea from the ITU document shown in Fig.~\ref{fig:EffGndCond} is too imprecise for use in a precise signal strength simulation. Furthermore, this document does not contain data for the area of North Korea. 

Another factor to consider when estimating SNR is noise strength. The major noise that affects the Loran signal is atmospheric noise originating from lightning, which has a peak at 10 kHz and a bandwidth between 1 kHz and 20 MHz \cite{BoyceJr07}. The Loran signal is a 100 kHz radio wave and is included in the atmospheric-noise bandwidth. We have previously calculated the atmospheric noise over Korea based on the expected values of atmospheric radio noise data published by the ITU \cite{ITU13:Radio}. Atmospheric noise varies with time and season. Examples of figures depicting the atmospheric noise for each season are included in our previous publication \cite{Rhee13}. The detailed process for calculating the noise level is described in \cite{BoyceJr07}. 

Based on the estimated signal strength and noise strength at each user location, the SNR of a received Loran signal can be estimated. However, the estimation accuracy of the SNR from previous studies degrades due to the limited ground conductivity data over Korea.

\subsection{Measurement Error Simulation}
\label{sec:MeasErrSim}
The standard deviation of bias-removed Loran TOR measurements from a  transmitter $i$ (i.e., $\sigma_i$) is a function of the jitter of the transmitter (i.e., $J_i$) and the SNR of the received signals from the transmitter (i.e., ${S\!N\!R}_i$) as shown in (\ref{eqn:MeasVar}) \cite{Lo08}.
\begin{equation}
  \sigma_i^2 = J_i^2 + \frac{337.5^2}{N_\mathrm{pulses} \cdot {S\!N\!R}_i}
  \label{eqn:MeasVar}
\end{equation}
where $N_\mathrm{pulses}$ is the number of accumulated pulses of the Loran signal that is determined by the group repetition interval (GRI) of the corresponding Loran chain. The number $337.5$ was calculated from a reference measurement \cite{Lo08}. 
The transmitter jitter can be caused by thermal noise, bandwidth limitation, improper impedance termination, asymmetries in rise and fall times, cross-coupling, and so on \cite{Hancock:Jitter}.

It should be noted that the raw measurement from a Loran receiver is not time-of-arrival (TOA) but TOR. The relationship between TOA and TOR is as follows \cite{Son18:Novel}: 
\begin{equation}
  {T\!O\!A} = \Delta t_\mathrm{rx0} + n {G\!R\!I} + {T\!O\!R}
  \label{eqn:TOA}
\end{equation}
where $\Delta t_\mathrm{rx0}$ is the receiver power-on time measured from the Loran epoch (i.e., January 1, 1958), and $n$ is an unknown integer. Because $\Delta t_\mathrm{rx0}$, $n$, and GRI are constants, the standard deviation of TOA is the same as that of TOR.

The estimated SNRs from the previous subsection are utilized to obtain $\sigma_i^2$ in (1), but it is also necessary to use the correct transmitter jitter to accurately estimate $\sigma_i^2$, which is directly related to the Loran positioning accuracy. Previous studies assumed a fixed value of 6 m \cite{Lo08} or 4 m \cite{Hargreaves15} as the jitter of all transmitters, which causes errors in the positioning accuracy simulation because in reality each transmitter has a different jitter.

\subsection{Positioning Accuracy Simulation}
\label{sec:PosAccSim}

\begin{figure}
  \centering
  \includegraphics[width=0.6\linewidth]{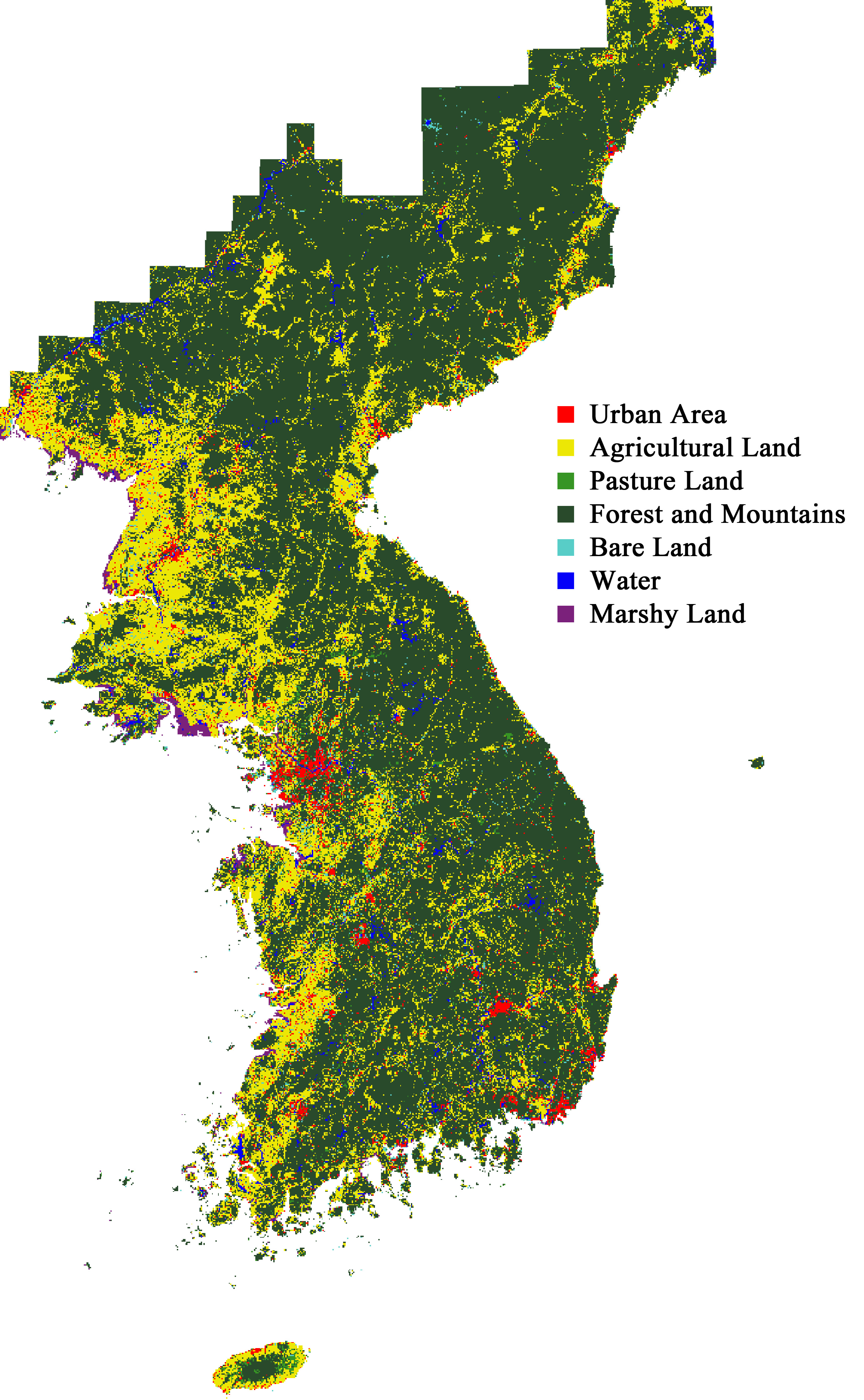}
  \caption{Land cover map from the Ministry of Environment, Korea \cite{EGIS:Republic}.}
  \label{fig:LandCover}
\end{figure}

Based on the estimated standard deviation of the bias-removed TOR measurements from transmitter $i$, which is $\sigma_i$, a weight matrix can be constructed as follows:
\begin{equation}
  W = \begin{bmatrix}
      \sigma_1^2 & \cdots & 0\\
      \vdots & \ddots & \vdots\\
      0 & \cdots & \sigma_N^2
      \end{bmatrix}^{-1}
  \label{eqn:WeightMat}
\end{equation}
where $N$ is the number of transmitters. With this weight matrix and a geometry matrix $G$, the position error covariance matrix can be calculated as in (\ref{eqn:CovarMat}) \cite{Enge06}. The geometry matrix $G$ contains the sine and cosine values of the azimuth angle $\theta_i$ from the user to the transmitter $i$.
\begin{equation}
\begin{split}
  \begin{bmatrix}
    \sigma_x^2 & \sigma_{xy} & \sigma_{xt}\\
    \sigma_{xy} & \sigma_y^2 & \sigma_{yt}\\
    \sigma_{xt} & \sigma_{yt} & \sigma_t^2
  \end{bmatrix}
  = (G^T WG)^{-1}\\
  \textrm{where} \quad G =
  \begin{bmatrix}
    \cos{\theta_1} & \sin{\theta_1} & 1\\
    \vdots & \vdots & \vdots\\
    \cos{\theta_N} & \sin{\theta_N} & 1
  \end{bmatrix}
\end{split}
  \label{eqn:CovarMat}
\end{equation}
This position error covariance matrix contains the covariance of each pair of random variables among $x$, $y$, and $t$, which represent the $x$-coordinate, $y$-coordinate, and user receiver clock. The $z$-coordinate is not considered because Loran provides only a two-dimensional position using low-frequency ground waves. 

Given this position error covariance matrix, a 95\% horizontal accuracy (i.e., two sigma) can be obtained using (\ref{eqn:Acc}).
\begin{equation}
  \mathrm{Accuracy} = 2 \sqrt{\sigma_x^2 + \sigma_y^2}
  \label{eqn:Acc}
\end{equation}
This accuracy assumes no bias in the position solution. In other words, the bias due to ASF is assumed to be compensated using spatial and temporal ASF correction methods \cite{Hargreaves12, Son18:Novel}. This type of accuracy is called a repeatable accuracy \cite{Peterson98}.

\section{PROPOSED METHODS TO IMPROVE THE SIMULATION ACCURACY}
\label{sec:ProposedMethods}

\subsection{Ground Conductivity Estimation from Land Cover Map}
\label{sec:LandCoverMap}
As mentioned in Section \ref{sec:SnrSim}, it is important to apply accurate ground conductivity data to properly simulate the received signal strength at a user location that is necessary for Loran accuracy simulation, as shown in the flowchart in Fig.~\ref{fig:FlowchartAccSim}. Although it is common to use the ground conductivity data from the ITU \cite{ITU92:World}, the rough ground conductivity data over Korea in Fig.~\ref{fig:EffGndCond} does not provide an accurate estimation of the received signal strength.

As an alternative, we utilized the relationship between the terrain types and their effective ground conductivity values, as shown in Table~\ref{table:EffGndCond} \cite{Barringer99:Radio}. To obtain fine-grid terrain-type data over Korea, we used the land cover map in Fig.~\ref{fig:LandCover} that was provided by the Ministry of Environment, Korea \cite{EGIS:Republic}. The land cover map contains information on land usage. We converted this land cover map to effective ground conductivity data using the relationship shown in Table~\ref{table:EffGndCond}. The resolution of the land cover map in Fig.~\ref{fig:LandCover} is far better than that of the ground conductivity map shown in Fig.~\ref{fig:EffGndCond}. The original resolution (i.e., the area covered by one pixel) of the land cover map was 30 m $\times$ 30 m, and we down-sampled the data to a 7 km $\times$ 7 km resolution to balance the simulation accuracy and computational time. Thus, we used approximately 1,600 ground conductivity values in the area over South Korea, which is significantly more than the 11 ground conductivity values of the 11 sectors in Fig. \ref{fig:EffGndCond}. The experimental validation of this approach and the improvement in the received signal strength simulation are discussed in Section \ref{sec:RSSSimResults}.

\begin{table}[t!]
\centering
\caption{Effective ground conductivity and relative dielectric constant values for various types of terrain (Table 2.1-2 of \cite{Barringer99:Radio}).}
\begin{tabular}{>{\centering\arraybackslash}m{4.2cm}|>{\centering\arraybackslash}m{1.5cm} >{\centering\arraybackslash}m{1.5cm}}
 \hline\hline
 Terrain type & Effective ground conductivity (S/m)	& Relative dielectric constant, $\epsilon$, (esu)\\
 \hline
 Seawater & 5 & 80\\
 Fresh water &	$8\times10^{-3}$ & 80\\
 Dry sandy, flat coastal land &	$8\times10^{-3}$ & 10\\
 Marshy, forested flat land &	$8\times10^{-3}$ & 12\\
 Rich agricultural land, low hills & $1\times10^{-2}$ & 15\\
 Pasture land, medium hills and forest &	$5\times10^{-3}$ & 13\\
 Rocky land, steep hills & $2\times10^{-3}$ & 10\\
 Mountainous & $1\times10^{-3}$ & 5\\
 Residential area &	$2\times10^{-3}$ & 5\\
 Industrial area & $1\times10^{-4}$ & 3\\
  \hline\hline
\end{tabular}
\label{table:EffGndCond}
\end{table}

\subsection{Transmitter Jitter Estimation}
\label{sec:JitterEst}
As shown in (\ref{eqn:MeasVar}), the jitter $J_i$ of transmitter $i$ is directly related to the standard deviation $\sigma_i$ of the bias-removed TOR measurements from the transmitter. Thus, it is important to use the correct jitter values to improve the Loran accuracy simulation performance. However, as discussed in Section \ref{sec:MeasErrSim}, a fixed value of 6 m or 4 m was previously assumed in the existing Loran simulators \cite{Lo08, Hargreaves15}.

\begin{figure}
  \centering
  \includegraphics[width=1.0\linewidth]{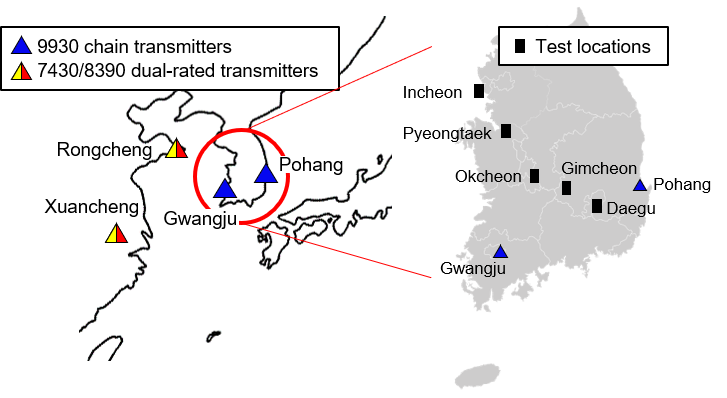}
  \caption{Loran chains and transmitters in Northeast Asia used in this study and our field-test locations in South Korea for evaluating the performance of the proposed received signal strength and measurement error simulation methods.}
  \label{fig:LoranChains}
\end{figure}

In this study, we propose a method to estimate the correct transmitter jitter based on actual TOR measurements, and the proposed method is experimentally validated. Four existing Loran transmitters in the Northeast Asia, whose signals were robust in Korea, were used in our accuracy simulator. The utilized transmitters were the Pohang (9930M) and Gwangju (9930W) transmitters in Korea and the Rongcheng (7430M) and Xuancheng (7430X) transmitters in China (Fig.~\ref{fig:LoranChains}).   

\begin{figure}
  \centering
  \includegraphics[width=0.9\linewidth]{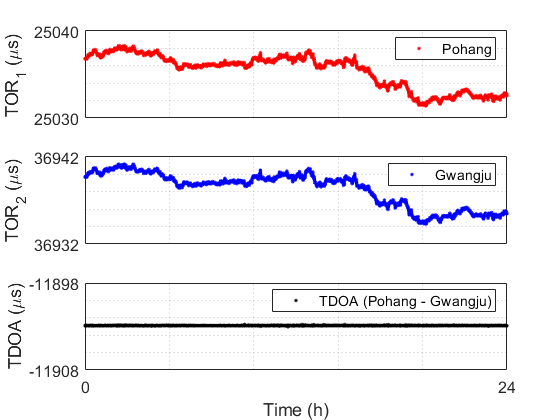}
  \caption{Example TOR measurements of the signals received from the Pohang (9930M) and Gwangju (9930W) transmitters and TDOA values between them over 24 hours in Incheon, Korea.}
  \label{fig:TORnTDOA}
\end{figure}

From (\ref{eqn:MeasVar}), the jitter $J_i$ of transmitter $i$ can be expressed as follows:
\begin{equation}
  J_i = \sqrt{\sigma_i^2 - \frac{337.5^2}{N_\mathrm{pulses} \cdot {S\!N\!R}}}
  \label{eqn:Jitter} 
\end{equation}
The TOR measurements from a Loran receiver contain certain biases that are not related to the transmitter jitter, such as the receiver clock bias, transmitter clock bias, and ASF variations. 
Thus, if the variance of the raw TOR measurements, $\sigma_{{T\!O\!R}_i}^2$, is used as $\sigma_i^2$ in (\ref{eqn:Jitter}), the estimated $J_i$ will also contain the same biases, which results in an incorrect transmitter jitter estimation. Therefore, it is important to remove the biases in the TOR measurements before calculating the variance $\sigma_i^2$. 

If the time-difference-of-arrival (TDOA) values between two transmitters are considered, the TDOA between transmitters 1 and 2 in the same Loran chain can be obtained as follows \cite{Son18:Novel}:
\begin{equation}
  {T\!D\!O\!A} = ({T\!O\!R}_1 - {T\!O\!R}_2) + ({E\!D}_1 - {E\!D}_2)
  \label{eqn:TDOA}
\end{equation}
where ${E\!D}_i$ is the known emission delay of transmitter $i$ with respect to the master transmitter of the same Loran chain.
The variance of TDOA, $\sigma_{T\!D\!O\!A}^2$, can be reliably estimated from raw TOR measurements because the common biases between ${T\!O\!R}_1$ and ${T\!O\!R}_2$ are canceled by (\ref{eqn:TDOA}). 

Because ${E\!D}_i$ is a known constant, the variance of the TDOA is expressed as follows:
\begin{equation} 
\label{eqn:VarTDOA}
\begin{split}
  \sigma_{T\!D\!O\!A}^2 
  & = \mathrm{Var} \left [ {T\!O\!R}_1 - {T\!O\!R}_2 \right ]  \\
  & = \sigma_1^2 + \sigma_2^2 \\
  & \; {  < } \; \sigma_{{T\!O\!R}_1}^2 + \sigma_{{T\!O\!R}_2}^2 
\end{split}
\end{equation}
where $\sigma_i^2$ and $\sigma_{{T\!O\!R}_i}^2$ are the variances of the bias-removed TOR measurements and the raw TOR measurements from transmitter $i$, respectively.
As shown in Fig.~\ref{fig:TORnTDOA}, the TOR measurements from the two transmitters contain significant time-varying biases, but it is noticeable that the common biases between $\textrm{TOR}_1$ and $\textrm{TOR}_2$ were eliminated in the TDOA values. Without eliminating the large biases in the TOR measurements in Fig.~\ref{fig:TORnTDOA}, $\sigma_{{T\!O\!R}_1}^2 + \sigma_{{T\!O\!R}_2}^2 = 3.8693 + 3.8727 = 7.7420 \, \mathrm{\mu s^2}$, which is orders of magnitude larger than $\sigma_{{T\!D\!O\!A}}^2 = 3.2977\times10^{-4} \, \mathrm{\mu s^2}$.
{  The time-varying biases dominate the value of $\sigma_{{T\!O\!R}_i}^2$, and 24-h data were used to calculate the sample variance as follows:
\begin{equation}
  \sigma_{{T\!O\!R}_i}^2 = \frac{\sum_{m=1}^{M} \left( {T\!O\!R}_i(m) - \overline{T\!O\!R}_i \right)^2}{M - 1}
  \label{eqn:DefSigTOR}
\end{equation}
where ${T\!O\!R}_i(m)$ is the \textit{m}-th raw TOR measurement from transmitter \textit{i}, $\overline{T\!O\!R}_i$ is the average of raw TOR measurements over the 24-h period, and \textit{M} is the total number of TOR measurements over the 24-h period.
}

To obtain $\sigma_i^2$, which is necessary for estimating the transmitter jitter, it is essential to remove the biases in the TOR measurements. We propose to detrend raw TOR measurements by applying Gaussian kernel smoothing to eliminate biases, which leaves only the random error components. As a metric of the bias elimination error, we suggest using:
\begin{equation}
  e = \lvert \sigma_1^2 + \sigma_2^2 - \sigma_\mathrm{TDOA}^2 \rvert.
  \label{eqn:ErrorMetric}
\end{equation}
Once TOR biases are properly removed, the sum of variances of bias-removed TOR measurements (i.e., $\sigma_1^2 + \sigma_2^2$) will be close to the variance of the TDOA measurements (i.e., $\sigma_{T\!D\!O\!A}^2$), as shown in (\ref{eqn:VarTDOA}), and the bias elimination error, $e$, will be close to zero. 

\begin{figure}
  \centering
  \includegraphics[width=0.9\linewidth]{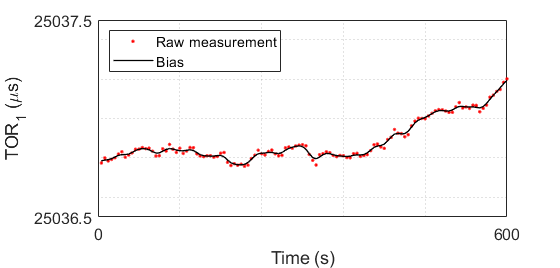}
  \caption{Bias in raw TOR measurements obtained by the Gaussian kernel smoothing. TOR measurements from the transmitter in Pohang are used in this example.}
  \label{fig:TORTrend}
\end{figure}

We estimated the TOR biases by smoothing the raw TOR measurements using a Gaussian kernel.
After applying the Gaussian kernel smoothing, the smoothed output signal ${T\!O\!R}_{i,\mathrm{bias}}$ from the biased raw input signal ${T\!O\!R}_{i}$ is obtained as follows \cite{Chung21:Gaussian}:
\begin{equation}
  {T\!O\!R}_{i,\mathrm{bias}}(t) = \int \! \frac{1}{b \sqrt{2\pi}} e^{-\frac{(t-s)^2}{2b^2}} {T\!O\!R}_{i}(s) \, ds
  \label{eqn:Gaussian}
\end{equation}
where $b$ is the bandwidth of the Gaussian kernel. Fig.~\ref{fig:TORTrend} shows an example of smoothed TOR outputs after Gaussian kernel smoothing, which represents the time-varying bias components in the raw TOR measurements. Once this bias is removed from the raw measurements, only random error components remain. Then, $\sigma_i^2$ can be calculated as:
\begin{equation}
  \sigma_i^2 = \mathrm{Var} \left[ {T\!O\!R}_i - {T\!O\!R}_{i, \mathrm{bias}} \right ].
  \label{eqn:Sigma_i}
\end{equation}

\begin{figure}
  \centering
  \includegraphics[width=0.8\linewidth]{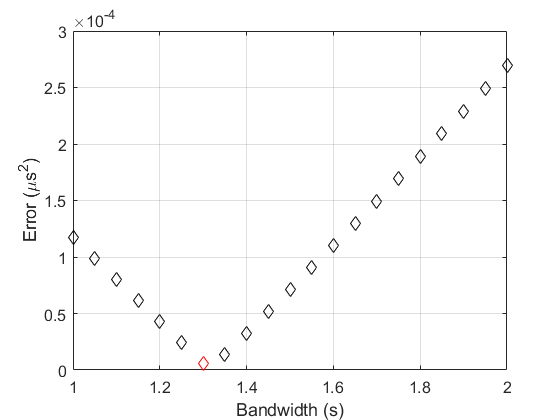}
  \caption{TOR bias elimination error according to the bandwidth of the Gaussian kernel. The optimal bandwidth in this example is $1.3$ s. TOR measurements from Pohang and Gwangju transmitters are used.}
  \label{fig:Error}
\end{figure}

\begin{figure}
  \centering
  \includegraphics[width=0.65\linewidth]{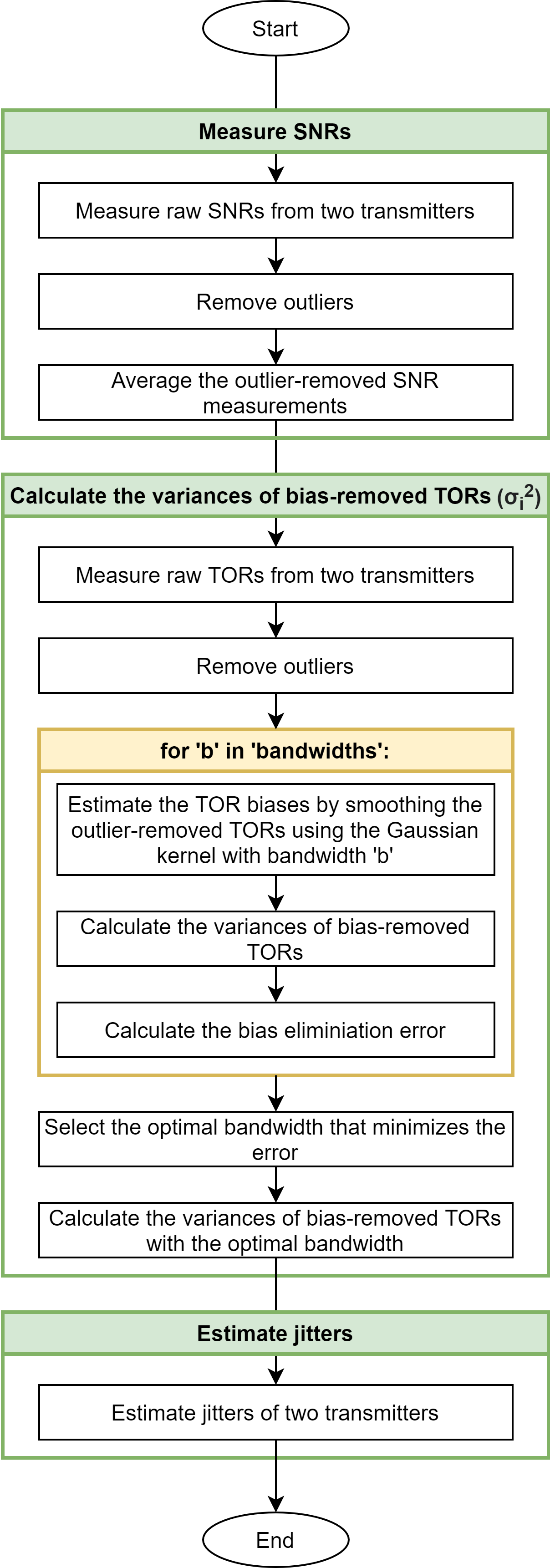}
  \caption{Flowchart of estimating Loran transmitter jitters. The jitters of two transmitters in the same Loran chain are estimated together.}
  \label{fig:FlowchartJitter}
\end{figure}

\begin{table*}
\centering
\caption{Comparison of signal strength (SS) [dB($ \mathrm{\mu}$V/m)] measurements and simulation results at five locations in Korea.}
\begin{tabular}{cccccccc}\toprule
\multicolumn{3}{c}{Test locations} & Incheon & Pyeongtaek & Okcheon & Gimcheon & Daegu \\
\midrule
\multirow{4}{*}[-7pt]{Pohang TX}
& \multicolumn{2}{c}{Measured SS} & 56.25 & 58.54 & 70.39 & 75.04 & 77.37 \\
\cmidrule(l){2-8}
& \multirow{2}{*}[-2.5pt]{Simulated SS}
& ITU map (existing) & 67.62	& 69.81 & 75.96 & 78.01 & 81.98 \\
\cmidrule(l){3-8}
& & Land cover map (proposed) & 65.19 &	66.97 & 71.87 & 75.85 & 81.26 \\
\cmidrule(l){2-8}
& \multicolumn{2}{c}{Improvement (\%)} & \textbf{21.37} & \textbf{25.25} & \textbf{67.71} & \textbf{72.72} & \textbf{15.51} \\
\midrule
\multirow{4}{*}[-7pt]{Gwangju TX}
& \multicolumn{2}{c}{Measured SS} & 57.45	& 57.50 & 61.32 & 59.56 & 57.08 \\
\cmidrule(l){2-8}
& \multirow{2}{*}[-2.5pt]{Simulated SS}
& ITU map (existing) & 64.56 & 66.99 & 70.76 & 69.67 & 68.19 \\
\cmidrule(l){3-8}
& & Land cover map (proposed) & 63.76 & 66.45 & 68.45 & 66.11 & 65.40 \\
\cmidrule(l){2-8}
& \multicolumn{2}{c}{Improvement (\%)} & \textbf{11.32} & \textbf{5.74} & \textbf{24.49} & \textbf{35.21} & \textbf{25.09} \\
\bottomrule
\end{tabular}
\label{table:RSSResult}
\end{table*}

It should be noted that the amount of smoothing depends on the bandwidth $b$ in (\ref{eqn:Gaussian}). Thus, $\sigma_i^2$ in (\ref{eqn:Sigma_i}) and the bias elimination error, $e$, in (\ref{eqn:ErrorMetric}) is also a function of $b$, which is explicitly expressed as:
\begin{equation}
  e(b) = \lvert \sigma_1^2(b) + \sigma_2^2(b) - \sigma_{T\!D\!O\!A}^2 \rvert.
  \label{eqn:ErrorMetricBW}
\end{equation}
Thus, the TOR bias elimination error can be calculated for each bandwidth selection, as shown in Fig.~\ref{fig:Error}. The optimal bandwidth is that which has a minimum error. In this case, $\sigma_1^2 + \sigma_2^2$ in (\ref{eqn:VarTDOA}) is closest to $\sigma_{T\!D\!O\!A}^2$, which indicates that the TOR biases are properly removed. With the optimal bandwidth, we calculated $\sigma_i^2$ of each transmitter and eventually obtained the transmitter jitter, $J_i$, from (\ref{eqn:Jitter}). 

The flowchart in Fig.~\ref{fig:FlowchartJitter} summarizes the proposed procedure for estimating the transmitter jitter. 
To remove outliers in the noisy SNR and TOR measurements, we used the outlier detection method suggested in \cite{Leys13}, which is based on the median absolute deviation (MAD). If a measurement was beyond the three-MAD range from the median of 100 recent measurements, it was identified as an outlier and removed \cite{Park20:Effect}.

\section{EXPERIMENTAL VALIDATION OF THE PROPOSED METHODS}
\label{sec:ExpValidation}

The proposed methods were validated through field tests, and the results are presented in this section.
As a metric of simulation accuracy improvement, we use the following formula:
\begin{equation}
\begin{split}
   \textrm{Improvement} \, (\%) &= \frac{E_\mathrm{existing} - E_\mathrm{proposed}}{E_\mathrm{existing}} \times 100 \\
   E_\mathrm{existing} &= \lvert {S\!R}_\mathrm{existing} - {GT} \rvert \\
   E_\mathrm{proposed} &= \lvert {S\!R}_\mathrm{proposed} - {GT} \rvert 
\end{split}
  \label{eqn:ImpMetric} 
\end{equation}
where $E_\mathrm{existing}$ and $E_\mathrm{proposed}$ represent the simulation error of the existing and proposed methods, respectively; ${S\!R}_\mathrm{existing}$ and ${S\!R}_\mathrm{proposed}$ represent the simulation results from the existing and proposed methods, respectively; and $GT$ is the ground truth value of the simulated variable.
Three aspects of the proposed simulation methods were validated: received signal strength, measurement error, and positioning accuracy simulations.

\subsection{Improvement on the Received Signal Strength Simulation}
\label{sec:RSSSimResults}

In Section \ref{sec:LandCoverMap}, we proposed the use of a land-cover map to indirectly obtain dense ground conductivity information.
To verify its effectiveness, the signal strength simulation results based on the ITU effective ground conductivity map (Fig. \ref{fig:EffGndCond}) and the land cover map (Fig. \ref{fig:LandCover}) were compared with the actual received signal strength measurements \cite{Kim20:Development} at five locations (i.e., Incheon, Pyeongtaek, Okcheon, Gimcheon, Daegu) across Korea, as shown in Fig. \ref{fig:LoranChains}. These locations between the Pohang and Gwangju transmitters were selected as they have geometric diversity across the country.
The Loran signals from the Pohang and Gwangju transmitters were received for 10--20 minutes at each location, and the average received signal strengths are presented in Table \ref{table:RSSResult}. 

At all five test locations, the proposed method to simulate the signal strength resulted in noticeable improvements over the existing method. The improvement measured by (\ref{eqn:ImpMetric}) varied between 5.74\% and 72.72\%. 
It is noteworthy that the proposed method provided lower signal strength estimations than the existing method for all the test cases. This is because the detailed ground conductivity information obtained from our method predicts additional signal attenuation that cannot be predicted by the coarse ITU ground conductivity data.

\begin{table}
\centering
\caption{Transmitter jitter [m] estimation results.}
\renewcommand{\arraystretch}{1.5}
\begin{tabular}{lccc||c}\hline
 Test locations & Okcheon & Gimcheon & Daegu & Average\\
 \hline
 Pohang TX & 2.67 & 1.75 & 1.90 & \textbf{2.11} \\
 Gwangju TX & 3.65 & 3.11 & 2.88 & \textbf{3.21} \\
 Rongcheng TX & 2.51 & 1.90 & 1.97 & \textbf{2.13} \\
 Xuancheng TX & 7.26 & 3.52 & 5.35 & \textbf{5.38} \\
 \hline
\end{tabular}
\label{table:EstimatedJitter}
\end{table}

\begin{table*}
\centering
\caption{Comparison of variance [$\mathrm{\mu s^2}$] of measurement errors at two locations in Korea.}
\begin{tabular}{ccccc}\toprule
\multicolumn{3}{c}{Test locations} & Incheon & Pyeongtaek \\
\midrule
\multirow{6}{*}[-14pt]{\shortstack{Pohang TX and \\ Gwangju TX}}
& \multicolumn{2}{c}{Measured $\sigma_{T\!D\!O\!A}^2$} & $3.2977\times10^{-4}$ & $2.7269\times10^{-4}$ \\
\cmidrule(l){2-5}
& \multirow{2}{*}[-10pt]{Simulated $\sigma_{1}^2 + \sigma_{2}^2$}
& Assumed 6-m jitter (existing) & $9.9163\times10^{-4}$ & $8.2276\times10^{-4}$ \\
\cmidrule(l){3-5}
& & Assumed 4-m jitter (existing) & $5.4657\times10^{-4}$ & $3.7770\times10^{-4}$ \\
\cmidrule(l){3-5}
& & Estimated jitters (proposed) & $3.5498\times10^{-4}$ & $1.8611\times10^{-4}$ \\
\cmidrule(l){2-5}
& \multicolumn{2}{c}{Improvement over the 6-m jitter case (\%)} & \textbf{96.19} & \textbf{84.26} \\
\cmidrule(l){2-5}
& \multicolumn{2}{c}{Improvement over the 4-m jitter case (\%)} & \textbf{88.37} & \textbf{17.55} \\
\midrule
\multirow{6}{*}[-14pt]{\shortstack{Rongcheng TX and \\ Xuancheng TX}}
& \multicolumn{2}{c}{Measured $\sigma_{T\!D\!O\!A}^2$} & $1.1630\times10^{-3}$ & $5.5246\times10^{-4}$ \\
\cmidrule(l){2-5}
& \multirow{2}{*}[-10pt]{Simulated $\sigma_{1}^2 + \sigma_{2}^2$}
& Assumed 6-m jitter (existing) & $1.4715\times10^{-3}$ & $8.2371\times10^{-4}$ \\
\cmidrule(l){3-5}
& & Assumed 4-m jitter (existing) & $1.0264\times10^{-3}$ & $3.7865\times10^{-4}$ \\
\cmidrule(l){3-5}
& & Estimated jitters (proposed) & $1.0426\times10^{-3}$ & $3.9481\times10^{-4}$ \\
\cmidrule(l){2-5}
& \multicolumn{2}{c}{Improvement over the 6-m jitter case (\%)} & \textbf{60.97} & \textbf{41.88} \\
\cmidrule(l){2-5}
& \multicolumn{2}{c}{Improvement over the 4-m jitter case (\%)} & \textbf{11.83} & \textbf{9.30} \\
\bottomrule
\end{tabular}
\label{table:MeasErrResult}
\end{table*}

\begin{table*}
\centering
\caption{Comparison of 95\% repeatable accuracy [m] at seven locations in Korea.}
\begin{tabular}{cccccccccc}\toprule
\multicolumn{3}{c}{Test locations} & Incheon & Pyeongtaek & Dangjin & Andong & Gumi & Jeonju & Gwangju \\
\midrule
\multirow{6}{*}[-36pt]{\shortstack{Pohang TX, \\ Gwangju TX, \\ Rongcheng TX, \\ and Xuancheng TX}}
& \multicolumn{2}{c}{Measured 95\% repeatable accuracy} & 10.16 & 8.72 & 10.09 & 12.73 & 8.87 & 8.49 & 12.13 \\
\cmidrule(l){2-10}
& \multirow{2}{*}[-10pt]{\shortstack{Simulated \\ 95\% repeatable \\ accuracy}}
& \shortstack{Assumed 6-m jitter with \\ the ITU map (existing)} & \multirow{1}{*}[4pt]{20.83} & \multirow{1}{*}[4pt]{18.19} & \multirow{1}{*}[4pt]{17.98} & \multirow{1}{*}[4pt]{22.41} & \multirow{1}{*}[4pt]{20.94} & \multirow{1}{*}[4pt]{15.36} & \multirow{1}{*}[4pt]{18.61} \\
\cmidrule(l){3-10}
& & \shortstack{Assumed 4-m jitter with \\ the ITU map (existing)} & \multirow{1}{*}[4pt]{14.04} & \multirow{1}{*}[4pt]{12.24} & \multirow{1}{*}[4pt]{12.11} & \multirow{1}{*}[4pt]{15.36} & \multirow{1}{*}[4pt]{14.28} & \multirow{1}{*}[4pt]{10.31} & \multirow{1}{*}[4pt]{12.84} \\
\cmidrule(l){3-10}
& & \shortstack{Estimated jitters with the \\ land cover map (proposed)} & \multirow{1}{*}[4pt]{12.10} & \multirow{1}{*}[4pt]{10.43} & \multirow{1}{*}[4pt]{8.67} & \multirow{1}{*}[4pt]{10.88} & \multirow{1}{*}[4pt]{9.86} & \multirow{1}{*}[4pt]{6.88} & \multirow{1}{*}[4pt]{11.50} \\
\cmidrule(l){2-10}
& \multicolumn{2}{c}{\shortstack{Improvement over the 6-m jitter with \\ the ITU map case (\%)}} & \multirow{1}{*}[4pt]{\textbf{81.82}} & \multirow{1}{*}[4pt]{\textbf{81.94}} & \multirow{1}{*}[4pt]{\textbf{82.05}} & \multirow{1}{*}[4pt]{\textbf{80.87}} & \multirow{1}{*}[4pt]{\textbf{91.82}} & \multirow{1}{*}[4pt]{\textbf{76.61}} & \multirow{1}{*}[4pt]{\textbf{90.25}} \\
\cmidrule(l){2-10}
& \multicolumn{2}{c}{\shortstack{Improvement over the 4-m jitter with \\ the ITU map case (\%)}} & \multirow{1}{*}[4pt]{\textbf{50.00}} & \multirow{1}{*}[4pt]{\textbf{51.42}} & \multirow{1}{*}[4pt]{\textbf{29.91}} & \multirow{1}{*}[4pt]{\textbf{29.61}} & \multirow{1}{*}[4pt]{\textbf{81.75}} & \multirow{1}{*}[4pt]{\textbf{11.68}} & \multirow{1}{*}[4pt]{\textbf{10.87}} \\
\bottomrule
\end{tabular}
\label{table:LornaPosAccSimResult}
\end{table*}

\subsection{Improvement on the Measurement Error Simulation}
\label{sec:MeasErrSimResults}

The jitters in the four transmitters in Fig. \ref{fig:LoranChains} were estimated using the method proposed in Section \ref{sec:JitterEst}. We used the TOR measurements of the four transmitters collected at three locations (i.e., Okcheon, Gimcheon, and Daegu) for the jitter estimation, and the estimated jitters are presented in Table \ref{table:EstimatedJitter}. Depending on the TOR measurement quality, the estimated jitters based on the measurements of one test location can be different from those of another test location. Thus, we estimated the jitters at three different locations and used the average value. 

The existing Loran performance simulators assume a fixed jitter value (e.g., 6 m \cite{Lo08} or 4 m \cite{Hargreaves15}) for all transmitters, as discussed in Section \ref{sec:MeasErrSim}. However, the estimated jitters were different between transmitters, as shown in Table \ref{table:EstimatedJitter}. 
Transmitter jitter is directly related to the variance of the bias-removed TOR measurements (i.e., $\sigma_i^2$), as in (\ref{eqn:MeasVar}). Thus, the effectiveness of the estimated jitters can be verified by comparing the measured and estimated variances using the jitters. However, it should be noted that the true value of $\sigma_i^2$ cannot be directly measured owing to the time-varying bias. 
Instead, the true values of $\sigma_{T\!D\!O\!A}^2$ were used as the ground truth for performance verification, as they can be directly measured because the common bias between two transmitters in the same Loran chain is automatically canceled.

If $\sigma_1^2$ and $\sigma_2^2$ of two transmitters in the same Loran chain are simulated with proper transmitter jitters, $\sigma_1^2 + \sigma_2^2$ should be close to the ground truth, i.e., $\sigma_{T\!D\!O\!A}^2$, as shown in (\ref{eqn:VarTDOA}). Table \ref{table:MeasErrResult} compares the simulation performance of the two existing methods (i.e., fixed jitters of 6 m or 4 m) and our proposed method (i.e., estimated jitters in Table \ref{table:EstimatedJitter}) at two locations (Incheon and Pyeongtaek).
The Pohang and Gwangju transmitters are considered together because they are in the same Loran chain as the Rongcheng and Xuancheng transmitters (see Fig. \ref{fig:LoranChains}).

Because the measurements collected at Incheon and Pyeongtaek were not used for the transmitter jitter estimation in Table \ref{table:EstimatedJitter}, they are suitable for verification purposes.
From among the five test locations in Fig. \ref{fig:LoranChains}, the data from three locations were used for jitter estimation, and the data from two locations were used for verification.  
Our estimated jitters clearly performed better than the fixed jitters at both test locations, as the simulated $\sigma_1^2 + \sigma_2^2$ was closer to the measured $\sigma_{T\!D\!O\!A}^2$ than were the fixed jitters. The performance improvement measured by (\ref{eqn:ImpMetric}) ranged from 9.30\% to 96.19\%, as presented in Table \ref{table:MeasErrResult}.

\subsection{Improvement on the Loran Positioning Accuracy Simulation}
\label{sec:PosAccSimResults}

\begin{figure}
  \centering
  \includegraphics[width=0.6\linewidth]{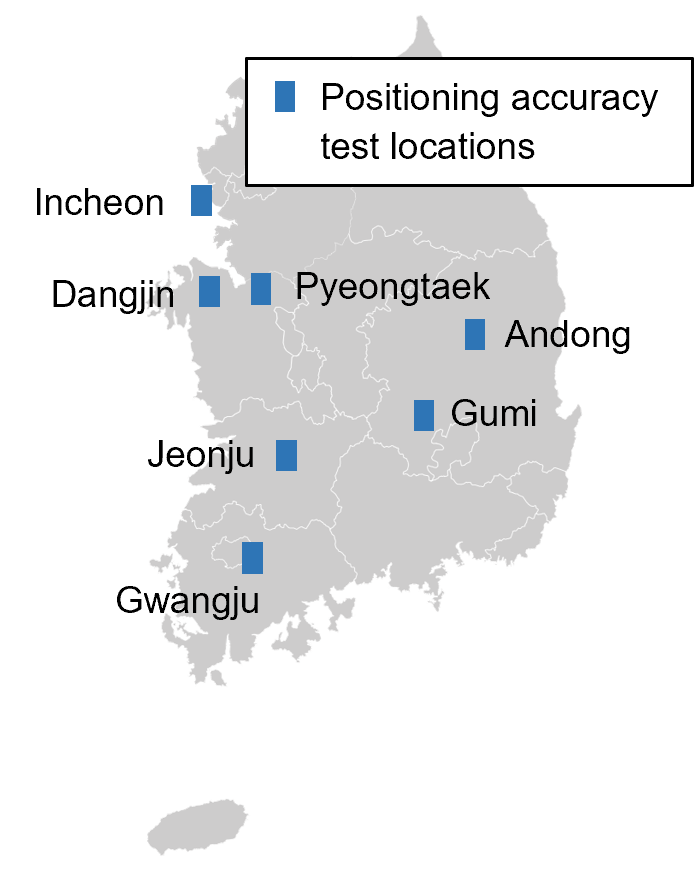}
  \caption{Field-test locations in South Korea for evaluating the performance of the proposed Loran positioning accuracy simulation method.}
  \label{fig:AccTestLocations}
\end{figure}

As explained in Fig. \ref{fig:FlowchartAccSim}, the simulated received signal strength in Section \ref{sec:RSSSimResults} and the measurement errors $\sigma_i^2$ in Section \ref{sec:MeasErrSimResults} were utilized to simulate the Loran positioning accuracy.
Because we have demonstrated that our methods provided better simulation accuracy of the received signal strength and $\sigma_i^2$, it is expected that the positioning accuracy simulation results from the proposed methods would also be better than those of the existing methods. This subsection verifies the comparison of the simulated and measured positioning accuracy results at the seven test locations in Fig. \ref{fig:AccTestLocations} over South Korea.

To calculate the measured positioning accuracy, TOR measurements from the four transmitters shown in Fig. \ref{fig:LoranChains} were used. Because the four transmitters belong to two Loran chains, the multichain Loran positioning algorithm \cite{Son18:Novel} was used to calculate Loran position solutions. To obtain a repeatable accuracy, spatial and temporal ASF errors were removed using the TDOA-based ASF correction method shown in \cite{Son18:Novel}. 
The 95\% repeatable accuracies at seven test locations in Incheon, Pyeongtaek, Dangjin, Andong, Gumi, Jeonju, and Gwangju were 10.16 m, 8.72 m, 10.09 m, 12.73 m, 8.87 m, 8.49 m, and 12.13 m, respectively, as shown in Table \ref{table:LornaPosAccSimResult}.
As in the case of Table \ref{table:MeasErrResult}, the Okcheon, Gimcheon, and Daegu data were not used for evaluating the positioning accuracy simulation performance, because the data from those sites were used for transmitter jitter estimation.
Our method using the estimated jitters with the land cover map demonstrated 10.87\% to 91.82\% improvement over the existing methods when the metric in (\ref{eqn:ImpMetric}) was applied.

\section{Conclusion}
\label{sec:Conclusion}
South Korea plans to proceed to a nationwide eLoran system to overcome the vulnerabilities of GNSSs after evaluating the eLoran performance in the testbed. To determine the optimal locations of future eLoran transmitters, it is essential to simulate expected positioning accuracy under a given transmitter distribution.
In this work, we proposed to utilize the land cover map to improve the quality of the  received signal strength estimation. A method to estimate transmitter jitters was also proposed to improve the quality of measurement noise estimation.
As a result, the performance of the Loran positioning accuracy simulation was improved by approximately 10\%--91\% compared with the existing methods, which was validated by the field-test data from seven locations across Korea. This improved Loran accuracy simulator is currently being used to provide the necessary information to decision makers in the Korean Ministry of Oceans and Fisheries regarding the future of Korea's nationwide eLoran system. Although the current focus of the Loran navigation performance in Korea is on the positioning accuracy, this Loran accuracy simulator can be expanded to simulate Loran integrity and availability in the future.

\section*{Acknowledgment}
The authors gratefully acknowledge the support of Sherman Lo of Stanford University for providing the Loran coverage availability simulation tool (LCAST), which was used as the baseline for this study.

\bibliographystyle{IEEEtran}
\bibliography{mybibliography, IUS_publications}

\phantomsection

\begin{IEEEbiography}[{\includegraphics[width=1in,height=1.25in,clip,keepaspectratio]{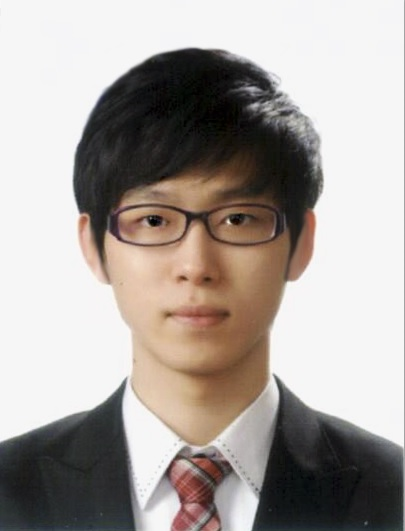}}]{Joon Hyo Rhee} received the B.S degree in electrical and electronic engineering from Yonsei University, Seoul, Korea, in 2012 and the Ph.D. degree in integrated technology from Yonsei University, Incheon, Korea in 2019. He is currently a senior researcher with the Korea Research Institute of Standards and Science, Daejeon, Korea. His research interests include complementary positioning, navigation and timing systems such as eLoran; intelligent transportation systems; and sensor fusion. Dr. Rhee was a recipient of the Graduate Fellowship from the ICT Consilience Creative Program supported by the Ministry of Science and ICT, Korea.
\end{IEEEbiography}

\begin{IEEEbiography}[{\includegraphics[width=1in,height=1.25in,clip,keepaspectratio]{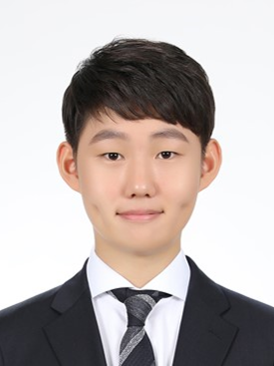}}]{Sanghyun Kim} received the B.S degree in integrated technology from Yonsei University, Incheon, Korea, in 2020. He is currently pursuing the Ph.D. degree in integrated technology at Yonsei University, Incheon, Korea. His research interests include complementary positioning, navigation and timing systems such as eLoran; intelligent transportation systems; and sensor fusion. Mr. Kim was a recipient of the Undergraduate Fellowship from the ICT Consilience Creative Program supported by the Ministry of Science and ICT, Korea.
\end{IEEEbiography}

\begin{IEEEbiography}[{\includegraphics[width=1in,height=1.25in,clip,keepaspectratio]{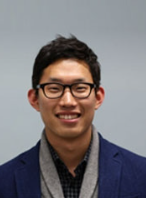}}]{Pyo-Woong Son} received the B.S degree in electrical and electronic engineering from Yonsei University, Seoul, Korea, in 2012 and the Ph.D. degree in integrated technology from Yonsei University, Incheon, Korea in 2019. He is currently a senior researcher with the Korea Research Institute of Ships and Ocean Engineering, Daejeon, Korea. His current research mainly focuses complementary PNT system including Loran. Dr. Son was a recipient of the Graduate Fellowship from the ICT Consilience Creative Program supported by the Ministry of Science and ICT, Korea.
\end{IEEEbiography}

\begin{IEEEbiography}[{\includegraphics[width=1in,height=1.25in,clip,keepaspectratio]{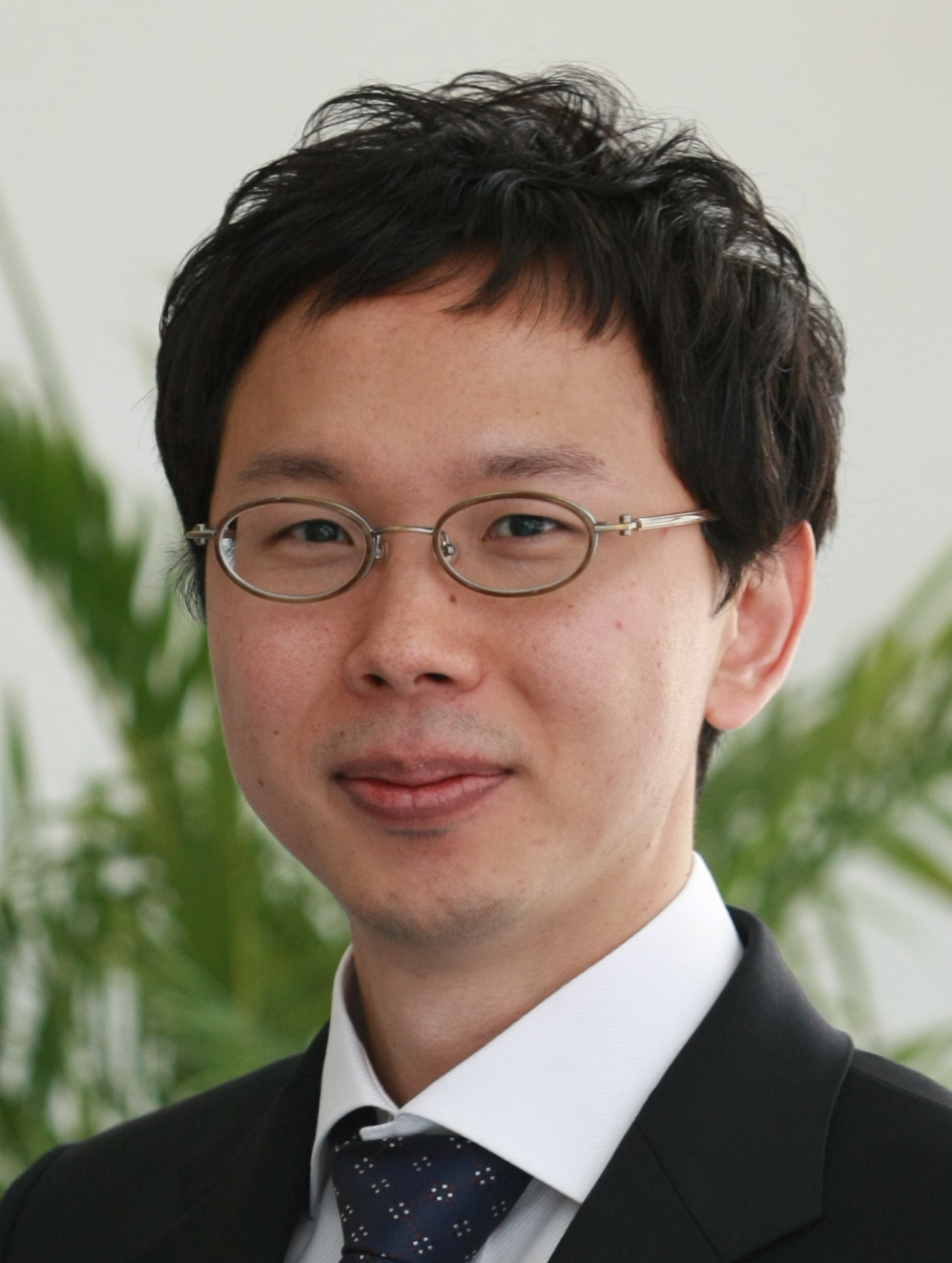}}]{Jiwon Seo} (M'13) received the B.S. degree in mechanical engineering (division of aerospace engineering) in 2002 from Korea Advanced Institute of Science and Technology, Daejeon, Korea, and the M.S. degree in aeronautics and astronautics in 2004, the M.S. degree in electrical engineering in 2008, and the Ph.D. degree in aeronautics and astronautics in 2010 from Stanford University, Stanford, CA, USA. He is currently an associate professor with the School of Integrated Technology, Yonsei University, Incheon, Korea. His research interests include GNSS anti-jamming technologies, complementary PNT systems, and intelligent unmanned systems. Prof. Seo is a member of the International Advisory Council of the Resilient Navigation and Timing Foundation, Alexandria, VA, USA, and a member of several advisory committees of the Ministry of Oceans and Fisheries and the Ministry of Land, Infrastructure and Transport, Korea.
\end{IEEEbiography}

\EOD

\vfill

\end{document}